\documentclass[preprint,superscriptaddress,floatfix,showpacs,balancelastpage,aps,pra,showkeys]{revtex4-2}
\usepackage[T1]{fontenc}
\usepackage[utf8]{inputenc}
\usepackage[UKenglish]{babel}

\usepackage{graphicx}

\usepackage{amssymb,amsmath}
\usepackage[hypertexnames=false]{hyperref}
\hypersetup{colorlinks=true,linkcolor=blue,citecolor=blue,urlcolor=blue}

\begin{document}

\title{Perspective on Quantum Bubbles in Microgravity}

\date{\today}

\author{Nathan Lundblad}
\affiliation{Department of Physics and Astronomy, Bates College, Lewiston, Maine 04240, US}

\author{David C.~Aveline}
\affiliation{Jet Propulsion Laboratory, California Institute of Technology,
4800 Oak Grove Dr.\ Pasadena, CA 90042, USA}

\author{Antun Bala\v{z}}
\affiliation{Institute of Physics Belgrade, University of Belgrade, Pregrevica 118, 11080 Belgrade, Serbia}

\author{Elliot Bentine}
\affiliation{Department of Physics, Clarendon Laboratory, 
Parks Road, Oxford, OX1 3PU, UK}

\author{Nicholas P. Bigelow}
\affiliation{Department of Physics and Astronomy, University of Rochester, Rochester, NY  14627  USA}

\author{Patrick Boegel}
\affiliation{Institut für Quantenphysik and Center for Integrated Quantum Science
and Technology (IQST), Universität Ulm, 89081 Ulm, Germany}

\author{Maxim A. Efremov}
\affiliation{German Aerospace Center (DLR), Institute of Quantum Technologies, 
Wilhelm-Runge-Straße 10, 89081 Ulm, Germany}

\author{Naceur Gaaloul}
\affiliation{Leibniz University of Hanover, Institute of Quantum Optics, 
Welfengarten 1, 30167 Hanover, Germany}

\author{Matthias Meister}
\affiliation{German Aerospace Center (DLR), Institute of Quantum Technologies, 
Wilhelm-Runge-Straße 10, 89081 Ulm, Germany}

\author{Maxim Olshanii}
\affiliation{Department of Physics, University of Massachusetts Boston, USA}

\author{Carlos A. R. S\'a de Melo}
\affiliation{School of Physics, Georgia Institute of Technology,
Atlanta, GA 30332, USA}

\author{Andrea Tononi}
\affiliation{Université Paris-Saclay, CNRS, LPTMS, 91405 Orsay, France}

\author{Smitha Vishveshwara}
\affiliation{Dept.\ of Physics, University of Illinois at Urbana-Champaign, USA}

\author{Angela C. White}
\affiliation{Australian Research Council Centre of Excellence in Future Low-Energy Electronics Technologies, School of Mathematics and Physics, University of Queensland, St.\ Lucia, Queensland 4072, Australia}

\author{Alexander Wolf}
\affiliation{German Aerospace Center (DLR), Institute of Quantum Technologies, 
Wilhelm-Runge-Straße 10, 89081 Ulm, Germany}

\author{Barry M. Garraway}
\affiliation{Department of Physics \& Astronomy, University of Sussex,
	Falmer, Brighton, BN1 9QH, UK}

\begin{abstract}

Progress in understanding quantum systems has been driven by the exploration of the geometry, topology, and dimensionality of ultracold atomic systems.  The NASA Cold Atom Laboratory (CAL) aboard the International Space Station has enabled the study of ultracold atomic bubbles, a terrestrially-inaccessible topology.  Proof-of-principle bubble experiments have been performed on CAL with an rf-dressing technique; an alternate technique (dual-species interaction-driven bubbles) has also been proposed.  Both techniques can drive discovery in the next decade of fundamental physics research in microgravity.

\end{abstract}


\maketitle


\section{Introduction}
\label{sec:introduction}

The study of ultracold quantum systems is a frontier in physics that has grown in an astonishing fashion in the twenty-seven years since the first observation of Bose-Einstein condensation (BEC) in 1995.  With a well-developed toolbox of forces used to confine, guide, and excite ultracold samples, physicists have used quantum gases to test fundamental ideas in quantum theory and statistical mechanics, as well as few- and many-body physics in general~\cite{Cooper.2019,giorgini:1215,Dalfovo:1999en,giorgini:1215,Fortagh:2007ji}. In particular, notions of geometry, topology, and dimensionality have directed the development of quantum-gas physics~\cite{Hadzibabic:2006lr,Clade.2009,Kinoshita.2004,Eckel:2014gf,Chin.2010,Blume_2012,RevModPhys.89.035006}. Quantum gases are typically confined in finite systems of some particular dimensionality and geometric character, such as a harmonic potential~\cite{jilabec}, a hard-walled “box”~\cite{Gaunt:2013ip}, or a periodic lattice potential~\cite{Greiner.2002}; looking beyond this, novel trapping geometries~\cite{PhysRevLett.86.1195} would permit (as it has in the past) the exploration of new realms of quantum physics and shed light on the nature of ultracold systems in general.

The \emph{bubble} geometry implies that the atoms, or especially a quantum gas, will move about a closed 2D surface. This differentiates a bubble from a droplet, which has a filled core.  The atoms, or superfluid in a bubble are trapped in a shell, which means that the simplest rotation of the fluid will produce two vortices at opposite poles. At present, the numbers of atoms involved typically range from many thousands to the million level, the temperatures are as low as tens of nano-Kelvin, and sizes vary from tens of micrometres to just under a millimetre. Experiments with ultra-cold quantum gases often use isotopes of rubidium, potassium, or sodium. These lend themselves to the production of quantum gases, such as Bose-Einstein condensates, although cold thermal gases can be of interest too. The bubbles themselves may be thick-shelled, or thin-shelled.

Bubble-shaped traps enabled by a microgravity environment offer a rich new geometry and topology in which quantum gas phenomena and related systems can be investigated. The collapse, expansion and excitations of such a quantum gas or BEC system are unexplored; the behaviour of quantum vortices in a BEC in an ultracold bubble has many associated open questions, and the nature and consequence of the crossover from a 3D (thick) shell to a quasi-2D (thin) bubble has stimulated significant recent theoretical work~\cite{Rhyno.2021,Moller:2020ik,Tononi:2022,Tononi.2022b,Tononi.2020,Tononi:2019ci,Bereta.2021lkc,Andriati.2021,Padavic:2020fj}.

In the following, we take a perspective view on the motivations, future developments and challenges involved in the creation and study of quantum bubbles.  In section~\ref{sec:motivation-cold-atom}, we summarise the general motivations behind the work in this area, and in sections \ref{sec:quantum-bubbles-rf} and~\ref{sec:quantum-bubbles-mixtures}, we explore two different approaches for the formation of quantum bubbles.  That is, in section~\ref{sec:quantum-bubbles-rf} we explore bubbles formed from shell potentials via radio-frequency dressed potentials. Then in section~\ref{sec:quantum-bubbles-mixtures} we explore bubbles formed on the surface of an inner droplet core made of different atoms where the physics of interactions of separated condensate mixtures is important.  Finally, we conclude with an outlook in section~\ref{sec:conclusion}.


\section{Motivation for cold atom bubbles in space}
\label{sec:motivation-cold-atom}

It is prohibitively difficult to generate ultracold bubbles in conventional terrestrial labs, due to gravitational sag.  With a harmonic trap gravitational sag will displace trapped atoms downwards. With a terrestrial shell trap the atoms will not be able to form a closed bubble unless the bubble is very small. To understand roughly the scale of the issue we can compare the gravitational potential energy difference `$mgh$' at the top and bottom of a shell to an ultracold thermal energy scale of $k_B T$. For rubidium atoms and a temperature of a micro-Kelvin the thermal energy corresponds to a height difference of about 10~$\mu$m. Although repulsive interactions can help with a condensate, this would be a very asymmetric bubble unless it was even smaller.  While in principle levitation techniques exist~\cite{Shibata:2020fx}, they lack the fine-tuning or uniformity to realistically generate bubbles in terrestrial labs.  A cold-atom machine in perpetual microgravity, such as the Cold Atom Lab (CAL)~\cite{Aveline:2020gk} or the Bose-Einstein Condensate and Cold Atom Laboratory (BECCAL)~\cite{Frye:2021dz}, is needed to explore this tantalizing scientific realm, although we note that terrestrial drop-and-catch machines, and apparatus aboard aircraft/spacecraft in ballistic flight, could in principle explore these phenomena~\cite{Condon:2019jk,Muntinga:2013ge,Barrett:2016ko,Lachmann.2021}.

\begin{figure}[t]
    \centering
    \includegraphics{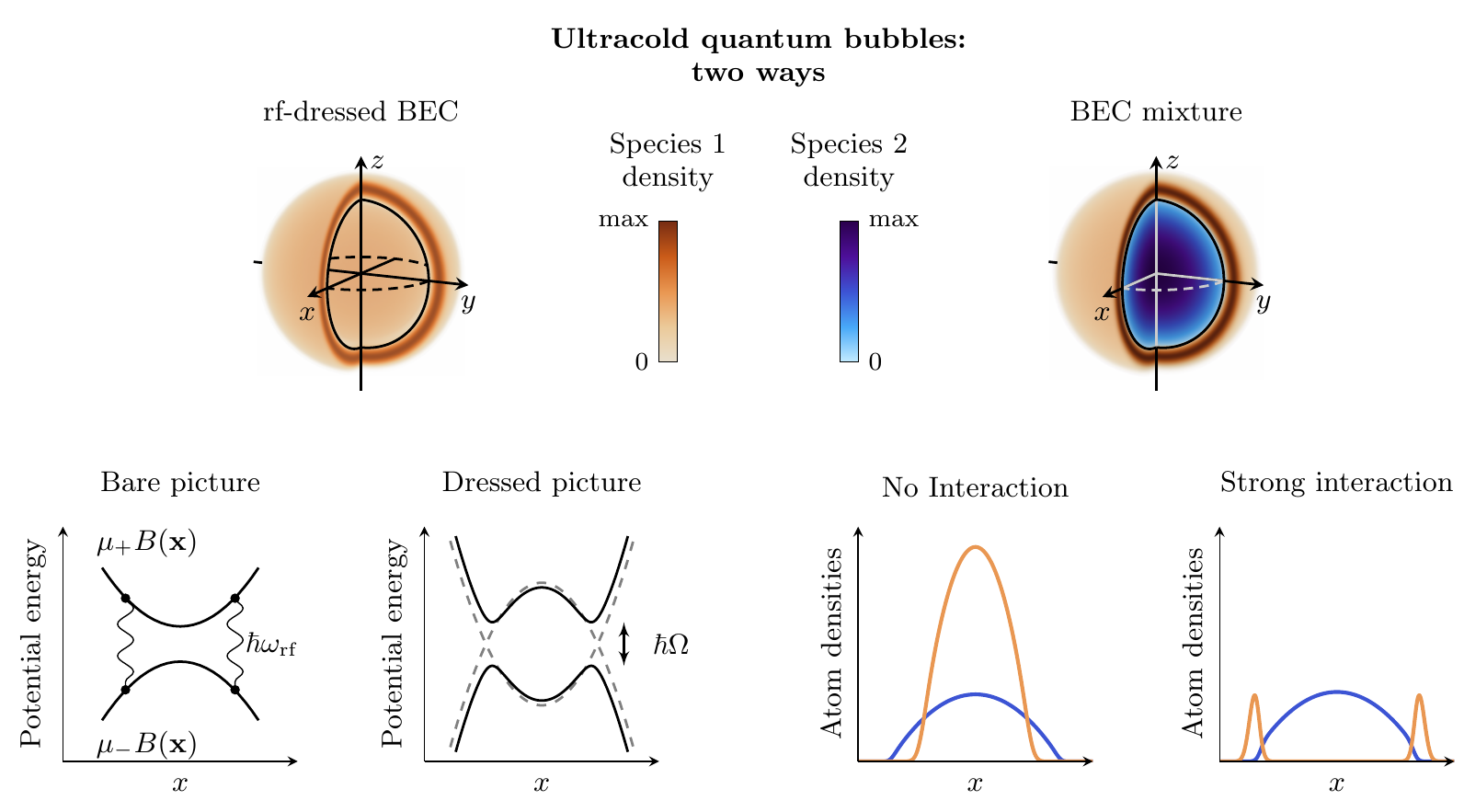}
    \caption{Two ways of realizing quantum bubbles: (upper row) Schematic sketch of 3D shell density profiles for the rf-dressed (upper left) and mixture (upper right) implementations. The darker colours illustrate higher atom density and the blue colour gradient illustrates the second species in the mixture case. (lower row) Underlying principle of shell generation. (lower left) Two-level system in a magnetic trap undergoing rf-dressing starting from the bare picture of a trapped and anti-trapped state with a resonant rf signal coupling the two states. The corresponding dressed-state picture depicts the double-well structure of the adiabatic potentials as seen in 1D which take the form of full bubbles in 3D. (lower right) Cuts through the ground-state density distribution of a dual-species mixture with and without inter-species interaction illustrating the hollowing of the outer species due to increased repulsive inter-species interaction.
    (Upper figure adapted with permission from Ref.~\cite{Wolf.2021}. Copyrighted by the American Physical Society.)
    }
    \label{fig:bubble_0}
\end{figure}

There are two different complementary techniques available to generate ultracold atomic bubbles in orbital microgravity: (i) radiofrequency (rf) dressing of ultracold atoms (see the reviews \cite{garraway2016review,perrin_advamop_2017}) and (ii) quantum gas mixtures with strong repulsive interactions (see e.g.~\cite{Wolf.2021}).  Figure~\ref{fig:bubble_0} illustrates the density profile in these two cases.  Exploring bubbles through both of these approaches will permit the precise investigation of both few- and many-body quantum physics emerging from the interplay of geometry, curvature, boundaries, and overall topology. The behaviour of solitons in a curved geometry could be explored, as well as the nature of the Kibble-Zurek mechanism in rapidly inflating (or deflating) bubbles~\cite{Bhardwaj.2021}.  Additionally, the physics of condensate shells are relevant to some astronomical models, including in neutron stars and other stellar bodies~\cite{Peralta.2005}.

Current theoretical work suggests that the hydrodynamic modes of an ultracold bubble should show a discontinuity at the transition from a filled system to the hollow shell~\cite{Sun:2018de,Padavic:2017cv,Wolf.2021}, further evolving into a two-dimensional bubble gas as the transverse confinement increases. Quasi-2D spherical shells should display a phase of Bose-Einstein condensation without superfluidity in the weakly-interacting finite-temperature regime~\cite{Tononi:2019ci}. The phase-space density and dimensionality of a harmonically trapped BEC changes dramatically as it is inflated to a bubble structure~\cite{Tononi.2020,Rhyno.2021}, thereby opening studies of finite-temperature effects in the quasi-2D regime such as the thermally-induced proliferation of vortices and Berezinskii-Kosterlitz-Thouless (BKT) physics~\cite{Tononi:2022,Mitra.2007,Mitra.2008}. The behaviour of quantum vortices on the surface of bubbles remains an area of open investigation \cite{caracanhas2022}, with key ideas including the constraints on superfluid flow imposed by topology, the potential creation of vortex lattices, and the lifetime of vortices in a system with periodic boundary conditions.

The long-term goal of this research program is to reach the threshold of understanding for the production, control and explication of the quantum physics of superfluids on curved manifolds, either open or closed.  This is of fundamental interest because quantum many-body systems have most often been studied in flat geometries, and characterizing the complex interplay of curvature, interactions and topology of a quantum gas will open new research directions in quantum systems generally.  For instance, controlling and understanding the curvature of a 2D quantum gas will enable interesting applications in quantum simulation. The existing BEC ring model of an expanding Universe \cite{PhysRevX.8.021021,Bhardwaj.2021} can be extended to the surface of an expanding sphere. There is scope for modelling aspects of planetary atmospheric dynamics by drawing on the analogue of quantum to classical turbulence on lengths scales where the nature of the vortex core is unimportant.  We can speculate that neutron stars could be simulated with bubbles. In any case, it is clear that with the local curvature constituting an additional tunable degree of freedom, there is potential to contribute to our understanding of fundamental physics: microgravity enables these developments.


\section{Quantum bubbles via radiofrequency-dressed potentials}
\label{sec:quantum-bubbles-rf}

The recent achievement of ultracold bubbles in orbital microgravity~\cite{Carollo.2021} proves that the International Space Station (ISS) is able to support cutting-edge investigations in this budding area (CAL was recently commissioned as an orbital BEC facility aboard the ISS~\cite{Aveline:2020gk}). This novel work was based on a microgravity implementation of long-standing theoretical proposals to use rf-dressed magnetic traps to create a bubble-shaped trap for ultracold atoms~\cite{PhysRevLett.86.1195,Lundblad:2019ia}. Such bubble traps are dynamically adjustable in their size and shape due to their dependence on common experimental quantities like trap currents and rf frequencies. While terrestrial shell structures have been observed~\cite{perrinbec,white:023616,Merloti:2013ft,Harte:2018jm}, they have always been radically deformed by the presence of terrestrial gravity, which causes the ultracold atoms to collect at the bottom of the bubble. This prevents their application to any physics requiring significant curvature or fully connected bubbles with cold atoms distributed over the entire surface.

\begin{figure}[htb]
    \centering
    \includegraphics[width=\columnwidth]{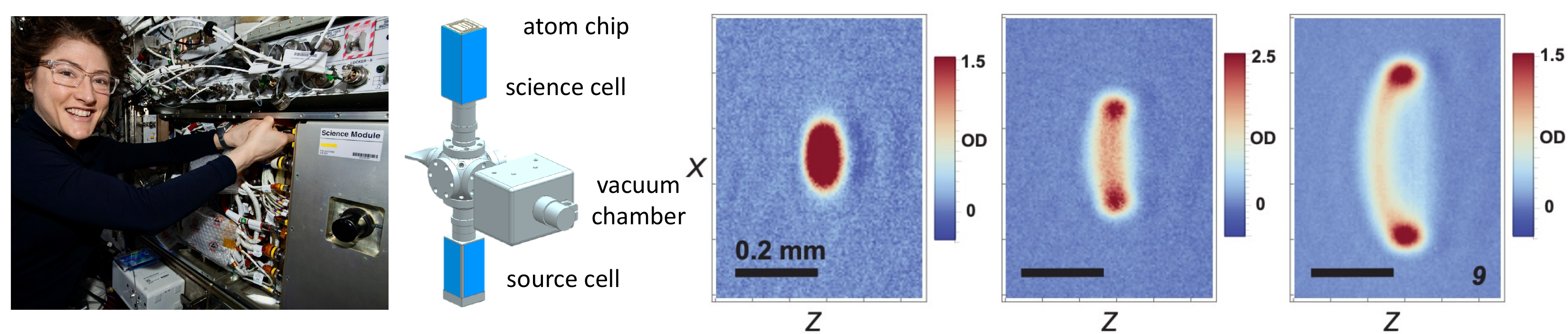}
    \caption{Recent observations of rf-dressed bubble structures in orbital microgravity with NASA CAL. Left photo shows astronaut Christina Koch installing a CAL upgrade on the ISS, and adjacent is an illustration of the vacuum chamber at the heart of the instrument.  Absorption image at left is of initial ultracold sample of $^{87}$Rb; the centre and right absorption images are of bubbles inflated to intermediate and large radii. The three absorption images portray false-colour optical depth, which (through column-density integration and significant imaging resolution effects) increases the apparent residual shell tilt and results in exaggerated `lobes', or a slight preference for higher densities at the top/bottom of the imaged bubble.  A terrestrial equivalent would have atoms strongly pinned to the leftmost 5\% of the bubble surface.  Figure taken and modified from Refs.~\cite{Aveline:2020gk,Carollo.2021}.}
    \label{fig:orbital}
\end{figure}

The rf-dressed approach to bubble creation boasts a demonstrated flight heritage as well as versatile tunability of shell size and thickness.  Rf-dressed traps have extremely low spurious heating rates as spontaneous emission is negligible, and they have very good lifetimes (approaching the vacuum-limited regime) when used with sufficiently strong coupling~\cite{Burrows:2017bw,lesanovsky:033619,Lesanovsky:2006fn}.  (In Ref.~\cite{garraway2016review} a 2~nK/s heating rate per RF antenna and terrestrial lifetimes of more than two minutes are reported.)  RF-dressed bubbles are also vulnerable to a variation in potential depth around the surface of the bubble, which arises from variations in the coupling strength of the rf field. These effects can be mitigated through careful rf coil modelling and design.  The technique is entirely insensitive to alignment as there are no elements to manipulate into position; this is helpful given the remote operation of CAL or similar facilities. The potentials are extremely smooth, standing out from other cold-atom technologies in this regard; optical potentials can suffer from wavelength-scale fringing imperfections, and even the residual roughness of static magnetic field traps is reduced by a smoothing effect associated with the rf-dressing~\cite{Hofferberth:2006p749}.

The underpinning rf technology allows considerable flexibility in controlling parameters of the shell and its deformations (when needed). Multiple frequencies can produce multiple concentric shells~\cite{Harte:2018jm}, the rf can be chirped or modulated to perform expansion, or breathing oscillations; rf polarization can be used for spatially varying surface effects \cite{Schumm_2005}, and the underlying magnetic field structure can be adjusted. Furthermore, in mixture experiments the bubbles for different species can be controlled in an independent manner~\cite{Bentine:2017ke}, offering a scheme to explore few-body physics and ultracold chemistry on curved manifolds. The bubble can be collapsed on demand by switching off, or adjusting the rf field strength, leading to bubble collapse physics, which is an unexplored avenue. Alternatively, the bubble shape can be conserved during the free evolution by applying matter-wave lensing techniques~\cite{Boegel2022}. In general, rf is a well-developed mature technology with a proven space track record, and as such it is sensible to rely on it as a key tool in this new domain of remotely-operated ultracold atomic-physics experiments. Future experiments in space would benefit through consideration of advancing these rf-dressing methods, as well as bubble generation through ultracold mixture interactions, discussed below.


\section{Quantum bubbles via ultracold mixture interactions}
\label{sec:quantum-bubbles-mixtures}

A complementary approach~\cite{Wolf.2021} for generating quantum bubbles in microgravity exploits optically trapped dual-species atomic mixtures and their tunable interaction. The conceptual idea is the following: The shape of the ground-state-density distribution of a mixture of two BECs~\cite{Pu:1998bj, Riboli.2002} strongly depends on the interplay between the intra-species and inter-species interaction. In an optical trap~\cite{grimmreview}, at least one interaction strength can be tuned with Feshbach resonances~\cite{Chin.2010} in a way that both species either spatially overlap, if the intra-species interactions dominate, or separate, if the repulsive interaction between the two species dominates over the intra-species interaction. In the immiscible regime and crucially in the absence of gravity, one atomic species is located at the centre of the trap while the other one forms a symmetric hollow shell around it, as displayed in Fig.~\ref{fig:bubble_0} (right). Such shell density distributions have been demonstrated in a terrestrial environment in two component condensates formed in magnetically trapped hyperfine spin states of $^{87}$Rb \cite{Hall1998, Mertes2007} and in optically trapped dual species BECs of sodium and rubidium \cite{Jia2022}. It is shown in Ref.~\cite{Wolf.2021} that the wave function of the outer shell of the mixture displays similar features to an rf-dressed shell BEC, in particular with respect to its ground-state density, collective excitation spectrum, and free expansion dynamics. Hence, the various studies of quantum few- and many-body effects on curved manifolds discussed above could also be performed with quantum bubbles realized with dual-species BECs.

In addition to the similarities with the rf-dressed scheme, the mixture approach~\cite{Wolf.2021} offers several experimental advantages or alternatives and unique applications that balance the technical challenges of dealing with a second species. For instance, the mixture can be directly condensed into a shell state without any need for adiabatic transition, which may improve the experimental cycle rate and reproducibility of the shell. Moreover, the mixture approach naturally provides ideal spherical symmetric configurations, where the optical potential of the trap is generated by a set of three identical orthogonal laser beams. In this way, fundamental physical effects on the shell could be studied more cleanly compared to ellipsoidal traps.

Another feature which naturally occurs in mixture-based shells, and not possible within rf-dressing technique, is the ability to dynamically magnify the shell~\cite{Wolf.2021} via its expansion. If the interaction between the two species is switched off after release from the trap, then the shell structure undergoes a drastic change and propagates both inwards and outwards resulting in an interference pattern, which occurs similarly in the rf-dressed case \cite{Lannert:2007kk,Tononi.2020}.  However, if the strong repulsion between the two species is kept on during free expansion, the spherically symmetric structure of the shell is conserved and both species expand simultaneously. This regime provides an easy way to magnify the dynamics on the shell and to better observe structures that are initially rather small, like vortex cores in the quantum gas. Furthermore, the possibility to tune the interactions of the mixture together with the usual mass imbalance between the two atomic species opens up new avenues for few body-physics \cite{Zhang:2018gl,Tononi.2022b} and molecule formation on curved manifolds. Finally, the response of the interacting mixture to the rotation of one of its components or the whole system would be fascinating to study from a fundamental point of view.


\section{Outlook and conclusion} 
\label{sec:conclusion}

Ongoing investigations in CAL's current science module (the `SM-3' generation) are focusing on reaching the Bose-condensed state in ultracold bubbles, having demonstrated in the first CAL science campaigns (`SM-2') that such bubbles are feasible.  Additional efforts are investigating mixtures of two species and applying alternative dressing techniques (including microwave transitions, or dual microwave/rf transitions) to broaden the scope of available configurations.  The atom-interferometer capabilities of CAL are also employed in the service of shell physics in the form of Bragg spectroscopy of dressed samples.  Ambitions for an upgraded CAL machine are aimed at initiation of vortex studies, potentially using a `stirring' tweezer beam or dynamic trap control of increased complexity; the elimination of all residual `accidental' inhomogeneities (as observed in figure~\ref{fig:orbital}) through careful field coil design would be ideal.

Aside from the ongoing progression of rf-dressing techniques within CAL's current capabilities, development of more advanced designs to achieve highly uniform and controllable bubbles would benefit from ground-based campaigns to mature essential technologies and define requirements for future space hardware. Generally, such hardware will need improved reliability and reduction of the size, weight, and power of ultra high vacuum chambers and pumps, as well as the many integral optical and electrical devices, such as lasers, fibre optics, and modulators. Specific development areas for quantum bubbles research include advanced RF and microwave sources, emitters, waveguides and/or cavities to improve coupling efficiency with trapped atoms and provide enhanced control of the dressing methods through precise control of rf polarization, frequency and power. To support shell experiments with mixtures, space-proven optical dipole traps and shielding capabilities for large magnetic fields should be developed further. In addition, extending the range of space-usable isotopes beyond rubidium and potassium would open up new opportunities for bubble mixtures.

We envision topical investigations with further CAL upgrades and follow-on instruments using both rf-dressing and inter-species interaction as bubble generators.  In the next decade, the NASA/DLR joint project BECCAL will operate aboard ISS and serve as a next-generation microgravity cold-atom machine, featuring design heritage from multiple terrestrial drop-tower and sounding-rocket architectures.  Following CAL upgrades and BECCAL, the use of cold-atom facilities in future space stations, or on dedicated free-flyer missions, could expand the scope and fidelity of the study of quantum bubbles and their topological analogues in fundamental physics research. We foresee quantum behaviour on curved manifolds with ultracold atomic bubbles---formed in at least two different ways as described here---becoming a featured physics `factory' of the orbital environment.  Such a factory would explore further questions: can dynamically-evolving bubbles give insight into cosmological Hubble physics~\cite{Banik.2021}?  Can we construct quantum-coherent bubbles at the millimetre scale, and what would their properties be?  Can the microgravity-enabled insights into the effects of topology and geometry be extended to new shapes like nested torii, M\"{o}bius strips, or lattice physics on bubbles?  The evolution from drop-tower, to sounding rocket, to CAL in orbit cements spaceborne cold-atom physics as a field ripe for exploration in the coming decade and beyond.

\begin{acknowledgments}
A portion of the CAL research was carried out at the Jet Propulsion Laboratory, California Institute of Technology, under a contract with the National Aeronautics and Space Administration (80NM0018D0004).  A part of this work was supported by the German Space Agency (DLR) with funds provided by the Federal Ministry for Economic Affairs and Climate Action (BMWK) due to an enactment of the German Bundestag under Grant Nos. 50WP1705 and 50WM2245B. The research of the IQST is financially supported by the Ministry of Science, Research and Arts Baden-Württemberg. A.B. acknowledges funding provided by the Institute of Physics Belgrade through the grant by the Ministry of Education, Science, and Technological Development of the Republic of Serbia. A. T. acknowledges support from ANR Grant Droplets No. ANR-19-CE30-0003-02.  Data statement: No new data were created or analysed in this study.
\end{acknowledgments}

%

\end{document}